# High contrast and resolution 3-D ultrasonography with a clinical linear transducer array scanned in a rotate-translate geometry.

**Théotim Lucas [1], Isabelle Quidu [2,3], S. Lori Bridal[1] and Jerome Gateau [1,*]**

[1] Sorbonne Université, CNRS, INSERM, Laboratoire d'Imagerie Biomédicale, LIB, F-75006, Paris, France; theotim.lucas@sorbonne-universite.fr (T.L.); lori.bridal@sorbonne-universite.fr (S.L.B.)
[2] CNRS, Lab STICC, ENSTA Bretagne, F-29806 Brest, France; isabelle.quidu@ensta-bretagne.fr
[3] Université de Bretagne Occidentale, F-29238 Brest, France
[*] Correspondence: jerome.gateau@sorbonne-universite.fr; Tel.: +33-144272265

**Abstract:** We propose a novel solution for volumetric ultrasound imaging using single-side access 3-D synthetic-aperture scanning of a clinical linear array. This solution is based on an advanced scanning geometry and a software-based ultrasound platform. The rotate-translate scanning scheme increases the elevation angular aperture by pivoting the array [-45° to 45°] around its array axis (axis along the row of its elements) and then, scans the imaged object for each pivoted angle by translating the array perpendicularly to the rotation axis. A theoretical basis is presented so that the angular and translational scan sampling periods can be best adjusted for any linear transducer array. We experimentally implemented scanning with a 5-MHz array. *In vitro* characterization was performed with phantoms designed to test resolution and contrast. Spatial resolution assessed based on the full-width half-maximum of images from isolated microspheres was increased by a factor 3 along the translational direction from a simple translation scan of the array. Moreover, the resolution is uniform over a cross-sectional area of 4.5 cm². Angular sampling periods were optimized and tapered to decrease the scan duration while maintaining image contrast (contrast at the center of a 5 mm cyst on the order of -26 dB for 4° angular period and a scan duration of 10 s for a 9cm³ volume). We demonstrate that superior 3-D US imaging can be obtained with a clinical array using our scanning strategy. This technique offers a promising and flexible alternative to development of costly matrix arrays toward the development of sensitive volumetric ultrasonography.

## 1. Introduction

Three-dimensional (3-D) ultrasound (US) imaging is increasingly prevalent in biomedical imaging [1]. Compared to conventional 2-D images, volumetric US display provides a detailed view of anatomical structures at various orientations. Thereby, visualisation of structures and the associated measurements are less dependent on the skill and experience of the sonographer and measurements can be more readily and independently repeated on the same data-set by other radiologists [2]. Additionally, volumetric US imaging facilitates fusion with images from other modalities for improved diagnosis or therapy planning. Furthermore, for longitudinal studies or for the monitoring of the progression or regression of pathology in response to therapy, 3-D US display facilitates the comparison of structures examined at different time points or by different practitioners.

Most current approaches developed to produce 3-D US images are based on volumetric reconstruction after acquiring successive 2-D images with an ultrasound 1-D array at different spatial positions and/or orientations [2]. For example, a 3-D ultrasound image of a fetus can be obtained using a mechanically-wobbled linear array [3]. After acquisition of a series of 2-D images, the volume is reconstructed from the 2-D envelope-detected images positioned in a 3-D volume using specific spatial transformations [4]. These transformations can be retrieved during free-hand scanning with position sensing [5,6], free-hand scanning without position sensing or mechanical scanning with a predefined motion [2]. Since these approaches use 2-D, envelope-detected images to produce a volumetric display, they can be implemented as external post-acquisition processes using images acquired with a conventional ultrasound machine. When a voxel has been viewed with different orientations of the array, incoherent spatial compounding can be applied to improve image quality [7]. However, the use of envelope-detected images limits the gain in terms of contrast or spatial resolution compared to what could be expected from coherent compounding [8]. To obtain the greater advantages provided by coherent compounding, ultrasound signals or images must be accessed prior to envelope detection.

At the cutting edge of ultrasound technology, planar arrays combined with programmable and fast acquisition electronics are emerging to achieve volumetric imaging rates in the kHz range. A variety of transducer spatial arrangements are proposed including row-column arrays [9], matrix [10,11] and sunflower [12] arrays. The ultrasound beam is electronically controlled and received signals are combined with coherent compounding. Synthetic aperture imaging [13,14], which consists in sequentially emitting and/or receiving with different array-element subsets, can be used when the number of array elements is larger than the number of transmit/receive channels. Planar array technology has attracted a lot of attention because it enables high-rate imaging of whole volumes and thereby enables advanced imaging modes throughout the volumes such as elastography [11], ultrafast Doppler [15] and super-resolution ultrasound imaging [16]. However, the technology is expensive since transducer planar arrays are still prototyping and a large channel count may be needed to probe the array aperture at a high rate. Additionally, the development of small but highly sensitive transducer elements that can be arranged in an array to cover a significant planar aperture with good spatial sampling is a complex technological challenge. Currently, planar arrays providing small elements and good spatial sampling sacrifice sensitivity and an overall aperture extent. Planar arrays with better sensitivity typically use larger elements at the expense of an element directivity (limiting the effective aperture) and degraded spatial sampling. Therefore, volumetric image quality with this technology has yet to rival that provided with 2-D array transducers.

Three-dimensional synthetic-aperture scanning provides an intermediate solution between the incoherent compounding of 2-D single-plane images and volumetric imaging with planar arrays. It consists in a mechanical scan of a transducer using a system providing access to the data prior to envelope detection, followed by coherent compounding of the signals acquired at different scan positions to synthetize a larger aperture during the volumetric image reconstruction process. The method has been developed for simple scanning geometries of clinical ultrasound arrays designed for 2-D imaging using: translation in the out-of-plane dimension [17], rotation around the median axis of the image [18] and a rocking motion [19]. The method advantageously uses already-developed transducer arrays that are technologically mature and highly sensitive, and facilitate the extension of methods to the clinical setting. The time required for mechanical displacements prohibits high-rate scanning of the volume, but the method provides higher image quality than the previously mentioned approaches because coherent compounding can be coupled with a high number of ultrasound signals recorded with very sensitive transducer elements. To date, 3-D synthetic aperture scanning has only been investigated in proof-of concept studies [17–19]. It has not yet been developed within a 3-D ultrasound scanner. One reason may be the relatively limited enhancement in spatial resolution that has been obtained with the simple scanning geometries that have been investigated thus far. Furthermore, ultrasound platforms that provide access to the raw channel or radiofrequency ultrasound data that is required for synthetic aperture scanning have only recently become widely available. Now is the time to reconsider 3-D synthetic aperture scanning with more advanced scanning geometries for applications that require high image quality but modest volume-scanning rates such as monitoring of a lesion's progression therapeutic response.

In this paper, we investigate the performance of a 3-D synthetic-aperture scanning scheme with a complex scanning geometry based on a rotate-translate scan with a linear ultrasound array. With the combined and predefined motions, a planar surface above the imaged volume is sequentially and efficiently populated with transducers having different spatial positions (array element and translated position) and orientations (angle of rotation about the normal to the surface). This combination of orientations provides a larger synthetic angular aperture in the elevation direction of the array, and thereby improves the resolution and contrast along the elevational axis. The proposed rotate-translate scanning approach was inspired from a scheme previously developed for photoacoustic (PA) tomography [20]. PA tomography is a hybrid imaging modality in which the ultrasound wavefield generated by laser excitation of optical absorbers is captured by transducer elements used only as detectors. The experimental implementation of the scanning scheme in this ultrasonic reception-only configuration resulted in high quality 3-D PA images. In the present study, we demonstrate for the first time the applicability of a rotate-translate scanning scheme for pulse-echo US volumetric imaging and we show experimentally that higher and more uniform 3-D resolution can be obtained than for existing synthetic aperture approaches. The original scheme proposed for photoacoustic imaging was limited due to the long scan duration. Herein, the scanning parameters have been optimized through theoretical considerations and experimental implementation to dramatically reduce the number of scanning positions and thereby the acquisition time while maintaining high contrast and uniform spatial resolution. In particular, we studied the influence of the rotation step on the spatial resolution and image contrast. The system was characterized *in vitro* with imaging phantoms and the performance was compared to synthetic aperture scanning with a translation-only scan to highlight the superior image quality of this new, more complex scanning scheme.

## 2. Materials and Methods

### 2.1. Presentation of the scanning scheme

The rotate-translate scanning scheme is diagrammed in 2-D (in the axial-elevation plane) for one element of the linear US array (**Figure 1**). The extension to 3-D is direct when the lateral dimension of the array (i.e. axis along the row of active array elements) is taken into account. We consider here a linear array with a cylindrical lens, which is typical of clinical linear US arrays for 2-D ultrasonography. The focusing power of the lens is usually weak which ensures a large depth-of-field but a relatively poor elevation resolution. These depth-of-field and resolution performances are directly linked to the small angular aperture $2.\beta$ of the element or large f-number = $F/D$ (**Figure 1** (a)) in the elevation direction. $F$ and $D$ are the focal distance and the width of the array element, respectively. To benefit from the large depth-of-field but with a significant improvement in the elevation spatial resolution, we developed a synthetic aperture approach. The synthesis of the aperture increases the angular aperture for each array element over an extended imaged area thanks to a rotate-translate scanning scheme. The rotation increases the angular aperture and the 1-D translation ensures coverage of the imaged zone.

To obtain a single-side access, the rotation axis was set to pass through the center of the transducer face $O_a$ (**Figure 1** (a)) and to be perpendicular to the axial direction of the transducer (**w**-axis) [20]. This axial direction is named transducer axis. The translation axis is perpendicular to both the rotation axis and the transducer axis when the rotation angle $\alpha_{rot}$ equals zero (**Figure 1** (b)). Translation across the imaged area is performed at a constant angle $\alpha_{rot}$, and then repeated for different angles of rotation (**Figure 1** (c)). All the translation scans have the same length $L$. To benefit from the full depth-of-field around the focus, the imaged area is centered on the focal distance $F$ in the $e_z$-direction (**Figure 1** (c)). As a consequence, the translational scan at each angle, $\alpha_{rot}$, is centered on the point $l_{trans\ center}(\alpha_{rot}) = -F^* \tan(\alpha_{rot})$ in the $e_x$-direction (**Figure 1** (c)). To have a constant step $\Delta l$ between successive translational positions of the transducer axis, the translational scan at $\alpha_{rot}$ is divided into tomographic positions separated by $\Delta l/\cos(\alpha_{rot})$.

The ultrasound pulse-echo signals are recorded for each tomographic position of the transducer element. In this configuration and with an adapted $\Delta l$ (see section 2.2), each point of the imaged area is probed with all the different orientations of the transducer. Therefore, for each pixel of the sampled area, the ultrasound signals from the different positions can be combined coherently during the reconstruction process to build a larger synthetic aperture. This process also corresponds to spatial coherent compound imaging in the elevation direction.

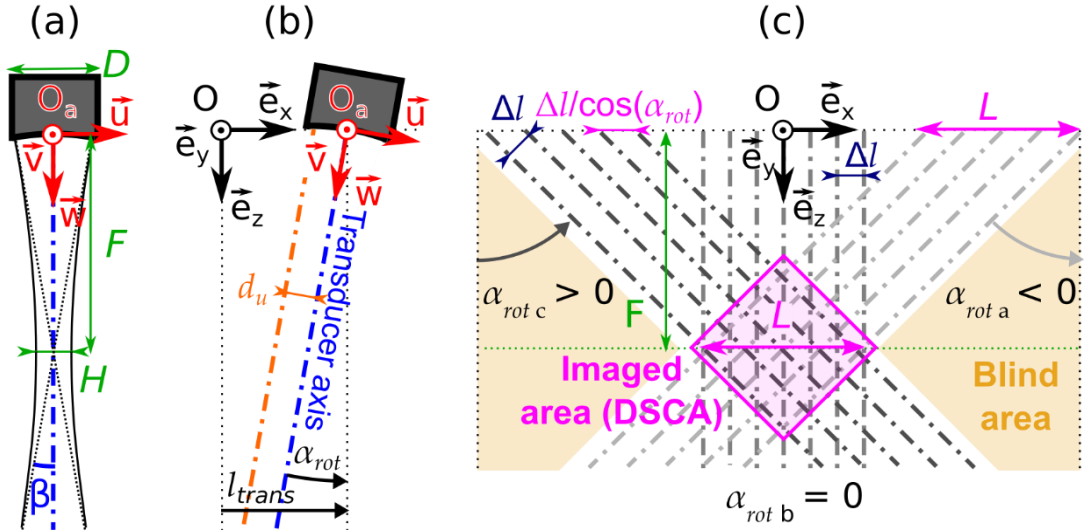

**Figure 1.** Schematic diagram of the scanning scheme in 2-D. (**a**) One weakly focused transducer array element of width $D$ and focal distance $F$. $H$ is the pulse-echo full width half maximum (FWHM) of the ultrasound beam. The angular full aperture is 2.$\beta$. The angle bisector is named transducer axis. A Cartesian coordinate system is attached to the transducer element face: **u** corresponds to the elevation direction and **w** to the axial direction. For a linear array, the long axis along the row of its active array elements (lateral direction) is the **v**-axis. The origin $O_a$ is the center of the transducer element face; (**b**) The transducer element at one tomographic position. For the fixed Cartesian coordinate system, the vectors $\mathbf{e_y}$ and $\mathbf{e_z}$ correspond to the rotation axis and the radial direction, respectively, when $\alpha_{rot}$ =0. The translational axis of the scan is $\mathbf{e_x}$. For the diagrammed position, the rotation angle is $\alpha_{rot}$ (here $\alpha_{rot}$ <0) and the translation position relative to the origin O is $l_{trans}$. $d_u$ is a distance to the transducer axis in the elevation direction used during the reconstruction process; (**c**) The transducer axis is shown for several positions of the scanning process. The translation positions are illustrated for three different rotation angles: $\alpha_{rot\ a}$, $\alpha_{rot\ b}$ and $\alpha_{rot\ c}$. For each angle $\alpha_{rot}$, the translation ranges from $l_{trans\ center}(\alpha_{rot})$ - L/2 to $l_{trans\ center}(\alpha_{rot})$ + L/2. For two successive positions at a constant $\alpha_{rot}$, the spacing between the transducer axes is $\Delta l$ and the transition step is $\Delta l/\cos(\alpha_{rot})$ along the translation axis. The magenta square indicates the imaged area, i.e. the region probed with all the different orientations of the transducer axis. This area is named DSCA for Diamond-Shaped Cross-sectional Area.

*2.2. Spatial sampling of the scan*

Our scanning scheme combines motion of the US transducer in translation and in rotation. Therefore, both the linear and the angular spatial sampling periods need to be set. To improve upon the approach proposed for PA tomography in [20], we sought to determine the sampling periods offering the best compromise between image quality and acquisition speed. Indeed, increasing the number of tomographic positions usually leads to higher image quality (contrast, homogeneity) but will result in longer acquisition time.

The minimal sampling periods required to avoid grating lobes and other aliasing artifacts on the image depend both on the transducer geometrical parameters (**Figure 1** (a)) and the reconstruction method. The reconstruction algorithm for 3-D imaging is detailed in Appendix A. Our reconstruction algorithm is based on a basic delay-and-sum process. In the 2-D configuration, the transducer is assumed to be a point transducer located at $O_a$ (**Figure 1** (a)), which emits circular waves (cylindrical in 3D) and has an isotropic detection pattern. The spatial impulse response of the transducer is not modeled, but its directivity is taken into account with an amplitude mask. For a given position of the transducer, the pulse-echo signal only contributes to the image in a slice of width $2*d_u$ (**Figure 1** (b)) around the transducer axis. In this slice, an apodization window (20% cosine taper) was used to avoid discontinuities. The pulse-echo (two-way) -12dB full width of the ultrasound beam is $1.2 * H$ [21] where $H = \lambda_c *$f-number, with $\lambda_c$ the wavelength at the center frequency $f_c$ of the transducer. So, to account for the focusing of the transducer, we set $d_u = 0.6 * H$.

The linear sampling period $\Delta l$ is defined as the distance between the transducer axis for two successive translation positions at a constant angle $\alpha_{rot}$ (**Figure 1** (c)). The angular sampling period, $\Delta \alpha$, corresponds to the angular change between two successive translations across the imaged area. Performing Nyquist–Shannon sampling in translation and in rotation [22], when considering point detectors in the reconstruction, would require a large number of positions and would lead to a prohibitive scan duration. To obtain more efficient scanning, we reduce the synthetic array to a rotation array: for each pixel in the imaged area, we coherently add signals corresponding a single position in translation for each rotation angle $\alpha_{rot}$. Thereby, we avoid generating a synthetic array with the successive translation positions at a given $\alpha_{rot}$. The selection of the translation positions is made based on the amplitude mask. We should have $\Delta l \geq 2*d_u$ so that $\Delta l \geq 1.2 *H$. On the other hand, an effective coverage of the imaged area would require $\Delta l \leq H$, because $H$ is the two-way full width half maximum (FWHM) of the ultrasound beam [23]. We chose to set $\Delta l = H$, as a compromise justified by the 20% cosine-taper of the apodization mask. One can note that $\Delta l$ and $d_u$ are strongly linked because of the reduction of the synthetic array to a rotation array. For the translation-only scan used for comparison, a synthetic translation array is formed. Therefore, the volumes probed by successive translation positions should overlap and then $\Delta l_{translation-only} \leq 2*d_u$. We set $\Delta l_{translation-only} = H/8$ [24] (see also Appendix B) while $d_u$ was still defined by $d_u = 0.6 * H$ to account for the transducer directivity.

Angular sampling in our scanning geometry has not yet been studied to the best of our knowledge. From studies performed for other geometries, we have found (Appendix B) that the minimal angular sampling period is $\Delta \alpha = \beta$. Herein, we study the influence of the spatial sampling on the imaged quality for 3 different nominal angular sampling periods:

$$\Delta \alpha_1 = \beta, \Delta \alpha_2 = \beta/2 \text{ and } \Delta \alpha_3 = \beta/4 \qquad (1)$$

$$\text{with } \beta = \text{atan}(D/(2*F)) \qquad (2)$$

We chose to perform our study with an angular range $\alpha_{rot} \in [-45°, 45°]$ which provides a large angular aperture for ultrasound imaging and will strongly showcase the potential of our approach. However, side lobes are expected from a finite and uniformly sampled angular aperture. They can be reduced with amplitude apodization whose amplitude value depends on $\alpha_{rot}$, but this solution leads to signal damping on the edge of the scanned interval and we consider this to be a sub-efficient use of the recorded signals. Therefore, we implemented spatial density tapering as a means to reduce both side lobes and the number of rotation positions. Because of its simplicity, we use the Doyle-Skolnik approach, as described in [25]. The angular positions of the synthetic array were determined considering a Hamming window function over the entire angular range [- 45°, 45°] and the nominal angular sampling period $\Delta \alpha$. The relative reduction in the number of rotation positions compared to a uniformly sampled array is approximately - 46 % (*i.e.* the area ratio between the Hamming and the rectangular windows). The translation endpoints for the different rotation positions are shown in **Figure 3** for a translation range $L = 30$ mm and the angular sampling period $\Delta \alpha_2$. Due to the spatial density taper, the effective angular sampling step is $\Delta \alpha_2$ around $\alpha_{rot} = 0$ and increases with the value of $|\alpha_{rot}|$. Therefore, the bounds of the angular range [- 45°, 45°] are not reached during the scan process. The effective bounds depend on the value of $\Delta \alpha$. With an amplitude apodization and the Hamming window function, the recorded signals at $\alpha_{rot} = \pm 45°$ would have had a weight of 0.08 compared to a weight of 1 around $\alpha_{rot} = 0$.

## 2.3. Experimental implementation

We experimentally implemented the scanning scheme presented above with a clinical linear ultrasound array. This section presents the tomographic ultrasound scanner, the practical implementation of the scan and the imaged samples.

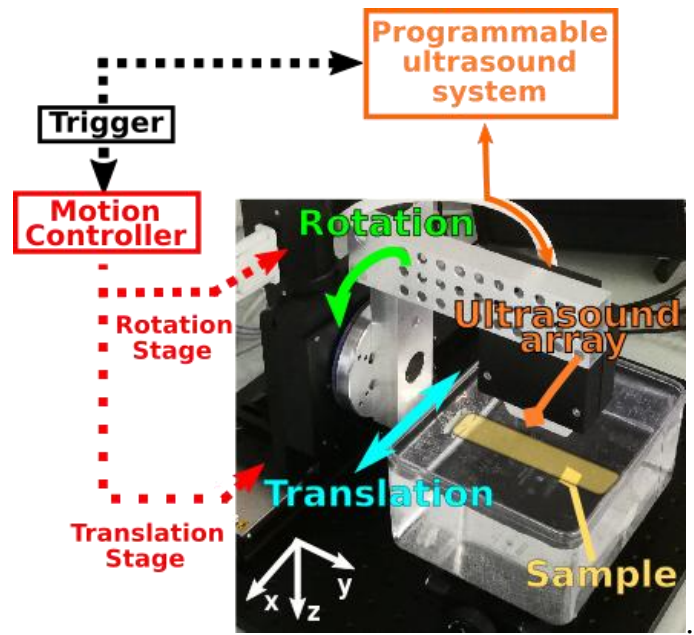

**Figure 2.** Annotated picture of the experimental setup. The rotate-translate scanner is comprised of two motorized stages that scan an ultrasound (US) array. The rotation stage is mounted on the translate stage. The US array is fixed on the rotation stage with custom-built holder. The array axis (axis along the row of its elements) was manually aligned to correspond to the rotation axis. Triggers synchronize 1) recording the motion-controlled stage positions and 2) the US acquisition with the programmable US system. The imaged sample is placed in the water tank below the array. The axes of the fixed coordinate system (O, $\mathbf{e}_x$, $\mathbf{e}_y$, $\mathbf{e}_z$) are represented. For the sake of readability, only the subscripts x, y and z are reported in the diagram. The y-axis and the z-axis correspond to the rotation axis and the radial direction when $\alpha_{rot} = 0$, respectively. Due to possible experimental misalignment, the translation axis is assumed to be close to the x-axis but may not exactly match. The scanner in action is shown in Video S1.

### 2.3.1. Experimental set-up

The experimental setup presented in **Figure 2** can be divided in three main parts: 1) the US acquisition system consisting of an US linear array driven by a programmable US platform, 2) the scanning system comprising two motorized stages and their motion controller and 3) the synchronization system piloted with a programmable trigger generator. The acquisition process was fully automated.

The ultrasound transducer array was a 128-element linear array (L7–4, Linear Array, Philips Medical Systems, Seattle, WA, USA) with a 5-MHz center frequency and 298-μm inter-element spacing. In the elevation direction, the elements of the array had an aperture of $D = 7.5$ mm and were cylindrically focused to a focal distance of $F = 25$ mm, resulting in a f-number of 3.3. Therefore, as described in equations (1) and (2), $\Delta\alpha_1 \approx 8°$, $\Delta\alpha_2 \approx 4°$ and $\Delta\alpha_3 \approx 2°$. Additionally, for this array, $\lambda_c = 0.3$ mm and $H = 1$mm with a speed of sound of 1500 m.s$^{-1}$. This linear array was chosen because it is widely used in the ultrasound community. Acoustic coupling between the sample and the ultrasound array was obtained by immersion in a water tank.

The array was driven by a programmable, 64-channel Vantage ultrasound system (Verasonics, Kirkland, WA, USA). For all the transmit events and all the receive events, only the 64 central elements of the array were used. Upon the arrival of each external trigger event, all these 64 array-elements were fired to produce a "plane wave" emission (beam that is unfocused in the lateral direction of the array or the **v**-direction shown in **Figure 1**) [26]. Transmitted pulses were one cycle long at 5.2 MHz. For most acquisitions, all the elements in the transducer array were fired at the same time to produce an untilted plane wave ($\gamma_{st}=0$) referred to as "straight transmit". We also implemented an emission mode in which tilted plane waves were generated with a different angles $\gamma_{st}$ for successive trigger events. The steering angles were controlled electronically by introducing a small delay between the firing of adjacent elements according to the desired inclination angle. In this mode named "steered transmit", a cyclic emission sequence was used and, for each cycle, the tilted plane waves were emitted with increasing angular steering (**Table 1**). The steered angles were chosen to ensure uncorrelated transmitted wave-vectors. Given the active aperture of the array and the wavelength $\lambda_c$ at its center frequency, the steered angles must theoretically be spaced by at least ~0.9° to have independent transmitted wave-vectors [8]. We chose here an angular spacing of 2°. Moreover, the steered angles were chosen centered around zero to match the directionality of the array elements.

Pulse-echo signals were recorded at a sampling frequency of 20 MS/s, with a constant gain (no time gain compensation was applied) and for a time interval adapted to the distance range between 5 mm and 60 mm in front of the array. The transmitted pulse amplitude and the reception gains were adjusted to ensure a good amplitude digitalization of received signals without saturation. Ultrasound signals were stored in internal memories of the acquisition system.

Table 1. Parameters of the different acquisition modes for a translation range of L=30 mm.

| Scanning mode | Rotate-translate scan | | | | Translate-only scan |
|---|---|---|---|---|---|
| | Straight transmit | | | Steered transmit | Straight transmit |
| Mode # | 1 | | | 2 | 3 |
| Angular sampling period | $\Delta\alpha_1 = \beta$ | $\Delta\alpha_2 = \beta/2$ | $\Delta\alpha_3 = \beta/4$ | $\Delta\alpha_2 = \beta/2$ | 0 |
| $\Delta\alpha$ (°) | 8 | 4 | 2 | 4 | 0 |
| Number of angular positions [1] | 6 | 12 | 23 | 12 | 1 |
| Steering angles $\gamma_{st}$ (°) | 0 | 0 | 0 | -4; -2; 0; 2; 4 | 0 |
| Acquisition rate (Hz) [2] | 50 | 50 | 50 | 250 | 50 |
| Number of tomographic positions [2] | 267 | 508 | 947 | 2533 | 244 |
| Acquisition duration (s) | 5.3 | 10.2 | 19 | 10.1 | 4.9 |
| Reconstruction duration (s) [3] | 11.2 | 20.9 | 39.7 | 95.2 | 14.2 |

[1] The number of angular positions results from the angular sampling period $\Delta\alpha$ and the density taper

[2] The acquisition rate corresponds to the pulse repetition rate (PRF) of the trigger generator, and the number of tomographic positions to the number of generated trigger pulses.

[3] The reconstruction durations were measured for the largest imaged volume $\Delta x*\Delta y*\Delta z = 30*19*30$ mm$^3$ and a pixel size of $px*py*pz = 72*144*72$ µm$^3$, resulting in total of 417*133*417 ≈ 23 million voxels. The volumetric image reconstruction was performed using a GPU unit NVIDIA GeForce RTX 2060 SUPER. For the modes 1 and 2, some tomographic positions were discarded for the reconstruction (see Appendix A).

We used two high precision motorized stages. The rotation stage (M-061.PD, Physik instrumente, Karlsruhe, Germany) was mounted on the translation stage (L-511.44AD00, Physik instrumente, Karlsruhe, Germany). The two stages were controlled with a C-884 DC Motor Controller (Physik instrumente, Karlsruhe, Germany). Using integrated encoders, the positions of the stages were recorded with an accuracy of 0.001 degrees and 50 nm, respectively, at the time of arrival of an external trigger pulse to the controller. The positions were stored in an internal memory of the controller.

To synchronize the US plane wave emission and the recording of the stage positions, a trigger generator (BNC Model 577, Berkeley Nucleonics, San Rafael, CA, USA) was used to send external triggers simultaneously to both the programmable US system and the motion controller (**Figure 2**). The number of pulses and the pulse repetition frequency (PRF) depend on the acquisition parameters (**Table 1**), and were programmed in the pulse generator prior to each acquisition.

For each tomographic acquisition, the following sequence of events was used. First, the motors were positioned at the first tomographic position and the ultrasound system was initialized. Then, the ultrasound system and the motion controller were set to wait for a trigger pulse. The first trigger pulse started the motion of the stages following programmed trajectories (see 2.3.2.). This trigger pulse and the next ones triggered the simultaneous recording of the stage positions and of the pulse-echo ultrasound signals. At the end of the scan, all the data from the ultrasound system and the motor controller were transferred to a computer.

### 2.3.2. Practical implementation of the scanning

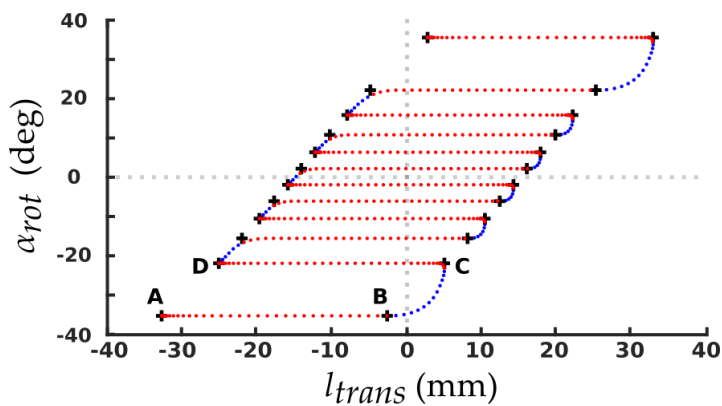

**Figure 3.** Successive motor positions for a rotate-translate scan with the angular sampling period $\Delta\alpha_2 \approx 4°$. Positions used for reconstruction are shown with red markers. Blue markers designate positions that are necessary for the continuous mechanical scan but are discarded prior to volume reconstruction. The black crosses represent targeted endpoints for the motor trajectories. Because of the density taper, the effective angular step is 4° around $\alpha_{rot} = 0$ and increases with the value of $|\alpha_{rot}|$. The bounds of the angular range [- 45°, 45°] are also not reached. The first four endpoints are notated A, B, C and D. The corresponding video of the scanner in action is displayed in Video S1.

Three different acquisition modes were employed. The first mode is a 3-D rotate-translate tomographic scan with straight transmit events ($\gamma_{st}=0$). The targeted endpoints of the motion are at $l_{trans\ endpoints}(\alpha_{rot}) = l_{trans\ center}(\alpha_{rot}) \pm L/2$ and are illustrated in **Figure 3** (black crosses). For a time-efficient scanning, we implemented continuous scanning rather than step-by-step scanning [27]. We chose to perform translation scans repetitively with a back and forth oscillatory motion. The rotation motion was performed in a single direction (increasing $\alpha_{rot}$). To illustrate the stage motion, we notated A, B, C and D the first four endpoints of the scan in **Figure 3**. For two successive rotation angles $\alpha_{rot\ 1}$ and $\alpha_{rot\ 2}$, the translation endpoints are scanned in one translation direction for $\alpha_{rot\ 1}$ (from A to B for instance) and in the reverse direction for $\alpha_{rot\ 2}$ (from C to D). Thereby, we minimized the scanning time. The rotation motion starts from the second translation endpoint at $\alpha_{rot\ 1}$ (B in the example) and we aimed at reaching the second rotational position $\alpha_{rot\ 2}$ when the targeted first translational endpoint at $\alpha_{rot\ 2}$ is reached (C here). A translation motion is additionally needed between the two endpoint positions (B and C in the example). In the example, the first, positive-direction translation occurs from A to C. The rotational motion starts when the translation stage reaches B. When C is reached, the rotation motion and the positive-direction translation end, and the translation direction is reversed.

The trajectories of positioning stages have trapezoidal or triangular velocity profiles. Therefore, for each movement, the following parameters had to be set: the endpoints, the targeted velocity and the acceleration. The deceleration was set equal to the acceleration. For the rotation stage, the endpoints $\alpha_{rot\,i}$ were defined by $\Delta\alpha$ and the spatial density tapper. The rotation velocity and acceleration were set to their maxima: 90 deg.s$^{-1}$ and 500 deg.s$^{-2}$, respectively. Most rotation moves had a triangular velocity profile. Their duration was then limited by the maximum rotation acceleration.

The translation velocity was set constant for each one-way translation at $\alpha_{rot}$ and equal to:

$$v(\alpha_{\rm rot}) = \frac{PRF}{N_{\rm st}} \frac{\Delta l}{\cos(\alpha_{\rm rot})}, \qquad (3)$$

where *PRF* corresponds to the trigger generator of the synchronization part, and $N_{st}$ is the number of electronic steering angles. $N_{st}$ = 1 in this acquisition mode as only straight transmits are performed. $\Delta l$ is equal to $H$ = 1 mm.

The translation acceleration and the *PRF* were determined empirically with the following criteria. We want that the rotation duration and the translation time between two successive endpoints at different $\alpha_{rot\,i}$ (such as B and C in **Figure 3**) are similar. The rotation duration is already reduced to its minimum, therefore we had to adapt the translation parameters. We decided to adjust the translation acceleration. We calculated the maximum translation acceleration so that the translation part is covered only when at least 80% of the corresponding rotation distance is traveled. With this acceleration, we determined the maximum translation velocity so that, for a translation travel equal to $L$ = 30 mm, 50% of the translation distance is traveled at a constant velocity (*i.e.* not in the acceleration or deceleration phase). The *PRF* was then deduced from equation (3) by taking the maximum possible angle $\alpha_{rot}$ = 45°. With these criteria, we found the following set of parameters: a *PRF* of $N_{st}$ * 50 Hz and, for the translation stage, an acceleration of 235 mm.s$^{-2}$ and a maximum velocity of 60 mm.s$^{-1}$ (maximum velocity from the manufacturer specifications: 90 mm.s$^{-1}$). With these parameters for $L$=30 mm and $\Delta\alpha_2$, 64% of the desired tomographic positions were acquired with an inter-position translation distance greater than 90% of $\Delta l/\cos(\alpha_{rot})$, the targeted translation sampling period.

The motion parameters (endpoints, direction and velocity) of both stages are coded in a macro, stored in the controller memory. Motion parameters are automatically modified each time an endpoint is reached. In particular, the controller starts the rotation motion when the translation coordinate of the corresponding endpoint is exceeded. This mode was implemented for the three angular sampling periods (**Table 1**).

The second acquisition mode is a 3-D rotate-translate tomographic scan with steered transmit events. The parameters for the stages are the same as for the first acquisition mode. However, we used $N_{st}$ = 5 steering angles and the PRF was thus increased to 250 Hz (**Table 1**). We can note that the translation velocities stay the same but the targeted inter-position translation distance is now $\Delta l/(\cos(\alpha_{rot})*N_{st})$. Consequently, the number of tomographic positions is around five times higher for the same $\Delta\alpha$. In this second acquisition mode, coherent compounding is performed both in the elevation direction of the array with the mechanical scanning and, additionally, in the lateral direction with electronic beam steering. We study the effect of the additional compounding on the contrast and resolution of the 3-D image. This mode was implemented only for the median angular sampling period $\Delta\alpha_2$.

The third acquisition mode is a translate-only scan with straight transmit events. The rotation stage was kept fixed at $\alpha_{rot}$ = 0, and a one-way translation scan was performed over the translation range $L$. The nominal translation sampling period $\Delta l_{translation-only}$ was set to $H/8$ = 125 μm [24] (see also Appendix B). In continuous motion with a PRF of 50 Hz, the corresponding translation speed was set to $v_0$ = 6.25 mm.s$^{-1}$. This third mode was implemented to provide a benchmark for comparison with the proposed rotate-translate scan.

2.3.3. Imaging phantoms

Four imaging phantoms were used to characterize our 3D scanner. The phantoms were totally submerged (**Figure 2**), maintained fixed compared to the water tank and centered in the imaged volume (**Figure 1** (c)).

The first phantom (Ph1) contains 200-μm diameter polyethylene microspheres (BKPMS 180-212 um, Cospheric, Santa, Barbara, CA) embedded in agar gel. The agar gel was prepared with agar powder 2% w/v (A1296, Sigma Aldrich) in water, and it was molded in 26.5-mm diameter cylindrical mold. The microspheres have a significant acoustic contrast relative to the embedding medium (on the order of 40 dB). They were randomly, but sparely, distributed over the entire cross section of the cylinder, and could thus be identified individually. Ph1 was placed so that all microspheres had approximately the same y-coordinate, i.e. they all lie within a plane perpendicular to the rotation axis. Ph1 was used to validate that the field of view can be adjusted by setting the translation range $L$ (**Figure 1** (c)). Moreover, Ph1 enabled us to study the 3-D spatial resolution and its spatial homogeneity.

The second phantom (Ph2) was prepared with agar powder (2% w/v) and 1% w/v cellulose powder (Sigmacell cellulose Type 20) in water. Cellulose particles (20 μm) act as ultrasound scatterers to mimic the scattering properties of biological tissues for US imaging. The gel was molded in the cylindrical mold with 3 cylindrical solid inclusions of 5 mm in diameter and of the same length as the mold. The inclusions were removed when the gel was solidified. Thereby, Ph2 contained 3 cylindrical holes arranged along the vertices of an equilateral triangle centered on the phantom axis. The center-to-center distance between adjacent holes was about 9 mm. The holes were filled with water and Ph2 was placed with the axes of the cylinders aligned with the rotation axis (y-axis). The holes are far enough from each other to be considered as isolated anechoic regions embedded in a speckle-generating medium. They were used to measure the image contrast and its spatial homogeneity.

Two other phantoms were made of 0.36 mm diameter polyester threads. The threads were verified to be acoustically scattering, however the contrast with agar gel was found to be poor. To obtain a higher contrast, the threads were simply placed in water: a non-scattering medium for 5 MHz ultrasound waves.

The third phantom (Ph3) is composed of four threads parallel to each other and oriented along the rotation axis (y-axis). Three of the threads are arranged along the vertices of an equilateral triangle of 9 mm side length. The fourth thread was at the center of the triangle. Ph3 allowed us to obtain an estimate of line spread function at four positions with a large dynamic range, and to derive a contrast metric named the cystic resolution. The similar spatial arrangement between Ph2 and Ph3 allow comparison of contrast measurements made with two suitable metrics (see section 2.5).

For the last phantom (Ph4), a 3-D net was formed with knots between twelve threads. The net is a complex structure comprised of a thin homogeneous scattering material. Because the threads are dyed in black, the ultrasound image of Ph4 can be easily compared with optical pictures. Ph4 allows us to study the ability to image a complex shape with multiple orientations.

*2.4. Image reconstruction algorithm*

Image reconstruction was performed with a simple delay-and-sum beamforming algorithm. The two-way travel times between the US transducer element positions and each imaged voxel were computed. Then, the value of each voxel was computed by summing, over all the elements of the array and all the different tomographic positions, the signal values assessed at the voxel-associated travel times. Tomographic positions outside of the translation limits of the theoretical scanning scheme (such as from B to C in **Figure 3**) were, however, discarded. A weighted sum was performed to implement several apodization windows and avoid side lobes and grating lobes, as detailed in Appendix A. Three-dimensional, envelope-detected images were obtained.

The 3-D display voxel grid is defined in the fixed coordinate system (O, $e_x$, $e_y$, $e_z$) presented in **Figure 1**(b). Voxel dimensions were chosen equal to $px*py*pz$ = 72*144*72 µm³, with an anisotropy reflecting the best-expected resolutions. The 3D images were displayed using maximum amplitude projection (MAP) images along the axes of the coordinate system. For instance, the MAP image along the x-axis was obtained by projecting the voxels with the maximum amplitude along the x-axis in the yz-plane. For each (y,z) coordinate, the pixel value of the MAP image corresponds to the highest voxel value along the x-axis of the full volume. When the MAP images are displayed with a logarithmic gray scale, the logarithmic compression is performed after the maximum amplitude projection.

## 2.5. Characterization of the imaging performance

We assessed the image quality in terms of resolution, contrast and spatial homogeneity using the phantoms presented in section 2.3.3. The full-width at half-maximum (FWHM) criterion is used to assess the 3-D resolution of individual point-like objects. For the contrast, two criteria were evaluated and compared: a variant of the cystic resolution (CyR) and the simple contrast between an anechoic region and a speckle region. We studied the influence of both the angular sampling parameter $\Delta\alpha$ and the transmit mode (straight or electronically steered).

### 2.5.1. FWHM resolutions

The 3-D resolution was assessed with phantom Ph1. To determine the FWHM resolution, the image of each microsphere was fit with a 3-D Gaussian blob using a nonlinear least-squares minimization. The three main axes of the blob were the axes of the Cartesian coordinate system, and the following equation was used:

$$g = A.exp\left\{-\left[\frac{(x-x_0)^2}{2.\sigma_x^2} + \frac{(y-y_0)^2}{2.\sigma_y^2} + \frac{(z-z_0)^2}{2.\sigma_z^2}\right]\right\}, \quad (4)$$

where $x_0$, $y_0$, and $z_0$ are the coordinates of the center of the blob, A is the amplitude, and $\sigma_x$, $\sigma_y$, and $\sigma_z$ are the Gaussian root mean square widths along the x, y, and z axes, respectively. The FWHM resolutions were computed using the equation: $FWHM_i = 2\sqrt{2ln(2)}\sigma_i$ with $i$ =x, y or z.

The expected FWHM resolution along the x-axis is on the order of $\lambda_c$, because of the large synthetic angular aperture. $\lambda_c$ = 0.3 mm is the wavelength at the center frequency of the array. Given the apodization along the y-axis (Appendix A), we expect a resolution on the order of 2*$\lambda_c$. The resolution along the z-axis is limited by the ultrasound pulse width and is expected to be on the order of $\lambda_c$. Accordingly, the fits with the Gaussian blob were performed in-volumes of 1.5 $\lambda_c$* 4 $\lambda_c$* 1.5 $\lambda_c$, each centered on an individual microsphere.

To compare the spatial resolution for different angular sampling periods, we used statistical quantifiers, such as the median and interquartile range (IQR), because of their robustness to outliers and reliability for various data distributions. The resolution comparisons are presented in boxplot displays.

### 2.5.2. The cystic resolution (CyR)

The CyR was introduced by Vilkomerson *et al* [28] to evaluate the effects of side lobes and grating lobes of the point-spread function (PSF) on the contrast of ultrasound images. The CyR is expressed as the minimum detectable size of anechoic structures embedded in a speckle-generating background. Instead of evaluating the detection of anechoic cysts of various sizes, Ranganathan *et al* [29] suggested using the image of the concentric PSF with a spherical void *c*. The CyR is then defined as the size of the spherical void that produces a contrast above a given threshold. In 2-D, the cyst contrast $Cc$ is computed [29] from the ratio of PSF energy outside the void relative to the total PSF energy:

$$Cc_{dB} = 10log_{10}\left(\frac{\iint_{(x,z)\notin c} h^2(x,z,x_0,z_0)dxdz}{\iint h^2(x,z,x_0,z_0)dxdz}\right), \tag{5}$$

where $h(x,z,x_0,z_0)$ is the image of a pointlike object centered at $(x_0,z_0)$, which stands for the PSF function.

For 3-D imaging, Rasmussen *et al.* [13] suggested to use the line spread function (LSF) instead of the PSF to access the CyR because the LSF is easier to measure experimentally and has a larger dynamic range. In our case, we are interested in measuring the effect of the scan parameters on the image quality. Therefore, we want to evaluate the CyR in xz-planes (perpendicularly to the rotation axis), using the LSF in the y-direction (along the rotation axis). Phantom Ph3 allows us to compute a robust estimate of the LSF with little background noise. The $Cc$ contrast was computed using the slice at y = 0 (center of the array) and the two closest uncorrelated xz-slices located at y = ±0.86mm. Each slice is a disc of 25*$\lambda_c$ diameter which is a compromise between the need to encompass the entire spatial extent of the LSF and the need to avoid overlap between neighboring LSFs. To obtain a smooth $Cc$ curve, the images were oversampled by a factor 2 in the xz-plane ($px = pz = 36$ μm). Because we expect anisotropy of the LSF, we extended the definition of the $Cc$ and we proposed a sectorial cyst contrast.

We can then introduce a general definition of $Cc$ as:

$$Cc_{dB}^T = 10log_{10}\left(\frac{E_T^{out}}{E_T^{tot}}\right), \tag{6}$$

$$E_T^{out} = \sum_{ny=-1}^{1} \iint_{(x,z)\in\{T\backslash c\}} h^2(x,y,z,x_0^y,z_0^y)dxdz, \tag{7}$$

$$E_T^{tot} = \sum_{ny=-1}^{1} \iint_{(x,z)\in T} h^2(x,y,z,x_0^y,z_0^y)dxdz, \tag{8}$$

where $T$ is either the entire slice (equation (6) is equivalent to (5) but for the LSF) or an angular sector of the slice (leading to sectorial contrast), $c$ is the disk-shaped void, and $T\backslash c$ is the set of elements of $T$ but not in $c$. The sectorial contrast was computed in angular sectors of $\pi/10$. $h(x,z,y,x_0^y,z_0^y)$ is the image of the LSF, at a given $y$, and centered at $(x_0^y,z_0^y)$, the coordinates of maximum amplitude of this LSF image.

Similarly to Rasmussen *et al.* [29], we define CyR as the cyst radius that produces a contrast $Cc_{dB}$ of -20 dB. For the entire disk-shaped slice, this definition of CyR is called isotropic cystic resolution, iCyR, in the following. For a sector, the sectorial cyst resolution is called anisotropic cystic resolution, aCyR.

2.5.3. Image contrast

Phantom Ph2 contains three anechoic structures in a speckle-generating medium that consist of cylindrical 5-mm diameter holes, invariant along the y-axis. To assess the detectability of this cyst size, we chose the simplest and most general contrast estimator used in ultrasound imaging [30]:

$$C0_{dB} = 10log_{10}\frac{\mu_c}{\mu_s} = 10log_{10}\frac{\frac{1}{N}\sum_{i=1}^{N}c_i^2}{\frac{1}{N}\sum_{i=1}^{N}s_i^2}, \tag{9}$$

where $\mu_c$ is the mean signal power in the cyst area and $\mu_s$ is the mean signal power in the speckled background area. The cyst statistics and the background statistics are assessed on the same number of pixels $N = 1000$, equivalent to about thirty resolution cells to guarantee reliable statistics. The variables $c_i$ and $s_i$ denote the value of $i^{th}$ pixel of the cyst area and of the background area, respectively. The background area was chosen distant from the three holes. As previously, the contrast is computed using the slice at y = 0 and the two closest uncorrelated slices (xz images). Each slice is a disc with a diameter limited to 25*$\lambda_c$.

As for estimation of the LSF based on $Cc$, a contrast curve was plotted by considering different and non-overlapping cyst areas of size $N$, all centered on the longitudinal axis of the cylindrical hole. The center-most area is a disk and the areas beyond this disc are concentric rings.

## 3. Results

### 3.1. Influence of the translation range on the imaged volume

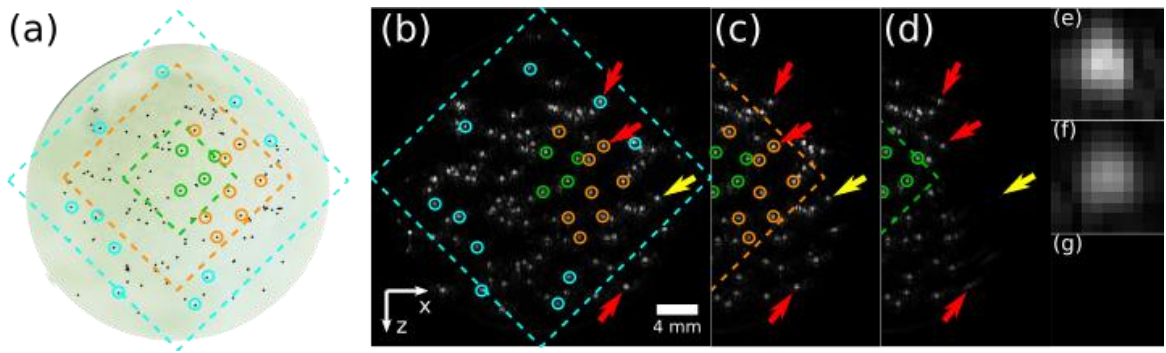

**Figure 4.** US images of Ph 1 performed with different translation ranges $L$. (**a**) Optical picture showing the spatial distribution of microspheres. (**b**) to (**d**) Maximum-amplitude projection (MAP) images along the y-axis for $L$ = 30, 20 and 10 mm, respectively. The dotted diamond indicates the area probed with all the different orientations of the array, named diamond-shaped cross-sectional area (DSCA) (**Figure 1** (c)). The MAP images are in linear grey scale. The same grey scale was used for (b)-(d). For a better readability, the grey scale was fixed. White pixels correspond to the maxima of the MAP image for L=30mm. Black pixels correspond to an amplitude equal to zero. The islets (e)-(g) show a zoom of the microsphere indicated with the yellow arrow for $L$ = 30, 20 and 10 mm, respectively. The pixels displayed in white correspond to the maximum for $L$ = 30 mm. The red arrows show microspheres located outside of the DSCA in (d). The yellow arrow points to a microsphere inside the blind area in (d). Full-size images corresponding to (c) and (d) are displayed in Figure S1.

The cross section of the volume probed at all orientations $\alpha_{rot}$ of the array and perpendicular to the rotation axis is presented in **Figure 1**(c). This cross-sectional area is expected to have a diamond shape and its area should depend on the translation range $L$. We name DSCA this diamond-shaped cross-sectional area. The cross section of the volume probed by none of the tomographic positions is named the blind area (**Figure 1** (c)). The regions outside of the DSCA (and the blind area) are probed with fewer orientations of the array. Therefore, we expect a degraded resolution and a lower amplitude for this part of the image.

To experimentally study the influence of $L$ on the cross section of the imaged volume perpendicular to the rotation axis, we performed several tomographic acquisitions of Ph1 for three different translation ranges. The angular sampling was kept constant and equal to $\Delta\alpha_2$. Ph1 contains 200μm-diameter microspheres that are sparsely distributed over the cross section of its 26.5-mm diameter cylindrical form (**Figure 4**(a)). We chose $L$ values equal to 30 mm, 20 mm and 10 mm, corresponding to 508, 396, 280 tomographic positions and scanning durations of 10.2, 7.9 and 5.6 s, respectively. The ratio of the scanning duration relative to $L$ becomes lower as $L$ increases, while the sampling parameters and in particular the translation velocities computed from equation (3) were identical for all scans. This is due to fact that the acceleration and deceleration phases of the translation motions have the same duration for all three scans, but these phases, during which the translation velocity is lower, are proportionally shorter as $L$ increases. The plane of microspheres was placed around y = 0 mm. Each microsphere could be identified individually allowing a qualitative evaluation of the effective imaged area and a comparison of the image quality inside and outside the DSCA. We reconstructed images of dimensions Δx * Δy * Δz = 30*2*30 mm³. **Figure 4** (b)-(d) presents the maximum amplitude projection (MAP) of the 3-D US image along the y-axis.

**Figure 4**(b) presents the US image for the largest translation range *L* = 30 mm. As illustrated by the circular markers which are landmarks, the spatial distributions of the microspheres in the US image and in the optical pictures match (**Figure 4** (a)). This result demonstrates the ability of our system to perform US images in the xz-plane and to localize small scatterers.

For *L*= 30mm, almost all microspheres are included in the DSCA. However, when the translation range is reduced, some microspheres enter the blind area on the side of the DSCA (along the x-direction), as illustrated by the microsphere marked with the yellow arrow in **Figure 4** which is no longer visible for *L* = 10 mm. Red arrows in **Figure 4**(d) illustrate that microspheres are indeed reconstructed with better resolution and amplitude when they are in the vicinity or inside the DSCA, compared to when they are located in the vicinity of the blind area. The islets (**Figure 4**(e-g)) further show the image amplitude degradation with a zoom on one microsphere.

With the developed scanner, the operator can adjust the translation range to fit the region-of-interest within the DSCA. Thus, the scan duration can be reduced when the object to image is smaller and the translation range can be more strongly limited.

*3.2. The FWHM resolution*

The resolution of isolated microspheres and their amplitude in the DSCA were evaluated for *L* = 30 mm according to the method detailed in section 2.5.1. Ph1 is comprised of about a hundred microspheres (**Figure 4**(a)). The final dataset excluded any pairs of microspheres positioned within 900μm of each other and contains 52 microspheres distributed across the DSCA (**Figure 5**(a)). Ph1 was also imaged using the translate only scanning mode (*L* = 30mm). For this mode, the resolution along the x-axis is degraded. The fits with the Gaussian blobs were performed in volumes of $6\lambda_c$ * $4\lambda_c$ * $1.5\lambda_c$ around each microsphere. Due to the small distance along x between neighboring microspheres, only 16 of them could be isolated in such volumes and included in the dataset. The microspheres have a finite diameter of about 200μm in diameter. Therefore, they cannot exactly be considered as point-like, and the measured FWHM values are expected to be slightly larger than the actual resolution of the system. However, this diameter of microspheres ensured an efficient scattering of the ultrasound waves and a sufficient contrast to the supporting medium (agar gel).

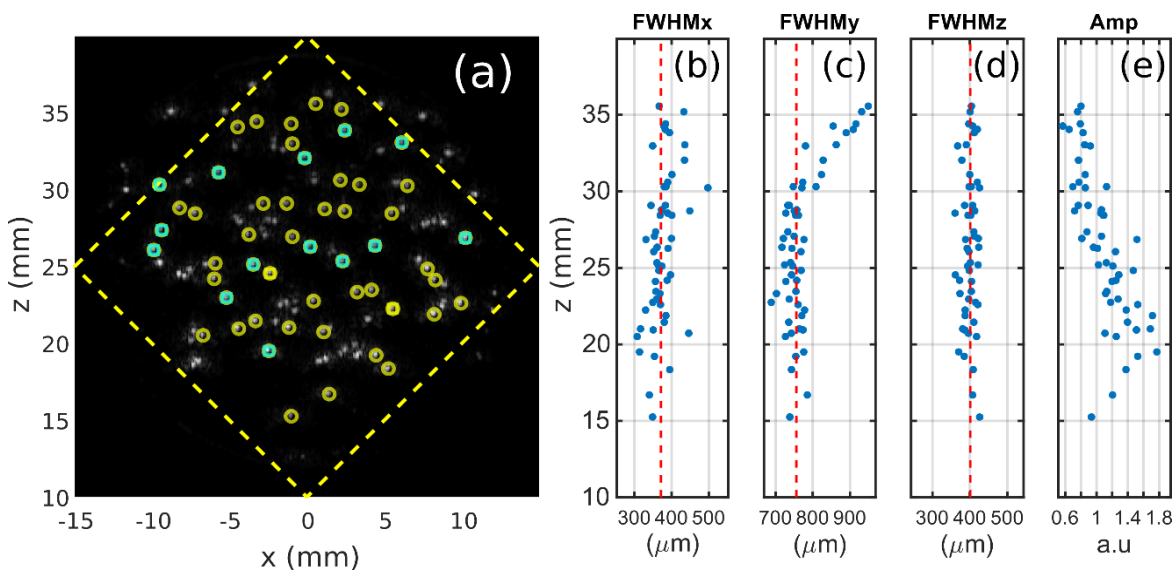

**Figure 5.** (**a**) MAP image along the y-axis of Ph1 for $\Delta\alpha_2$ in straight transmit. A linear gray scale was used. The 52 microspheres selected for the analysis are emphasized with yellow circular markers and the 16 microspheres kept for resolution estimations in translation-only mode are additionally

emphasized with square cyan markers. (**b**)-(**e**): Resolution and amplitude of the 52 microspheres in the image are shown as a function of the z-coordinate. The red-dotted lines indicate the median values. For each FHWM graph, the range of abscissa is equal to 300μm.

**Table 2** compares the median ± IQR values of the FWHM along the three directions of the coordinate system for the rotate-translate scan at $\Delta\alpha_2$ and the translate-only scan. We note that FWHMy and FWHMz are barely influenced by the rotation scan. Indeed, the resolution along the y-axis is mainly limited by the finite length of the active aperture of the linear array, and the resolution along the z-axis is mostly influenced by the duration of the pulsed echo. **Figure 5** (c) shows that FWHMy has a small deviation for z<30 mm due to the fixed lateral f-number used in the reconstruction method (Appendix A). For z>30 mm, the effective lateral f-number could no longer be maintained fixed by adjusting the apodized reception aperture of the array as the full length of the array is used. Therefore, the effective lateral f-number increases and the resolution degrades with the increasing depth z. The FWHMy values reported in **Table 2** were computed for the full range of depths. **Figure 5** (c) illustrates that the estimated median value of FWHMy is robust to the outliers for z>30 mm. The reported IQR values, which measure the statistical dispersion are, however, larger than they would have been if the microspheres at z> 30 mm had been discarded. **Figure 5** (d) shows the homogeneity of FWHMz with z. Unlike FWHMy and FWHMz, FWHMx was found to be 3.1 times smaller with the rotate-translate scan. The resolution along the x-axis is linked to the angular aperture along the elevation direction of the array. This angular aperture was increased from 2*β = 16° in the translation-only mode, to almost 90° (with a Hamming apodization) in the rotate-translate mode. Both FWHMy and FWHMx are limited by diffraction. However, **Figure 5** (b) demonstrates that FWHMx is noticeably more homogeneous even for z>30 mm, as opposed to FWHMy. Thus, our scanning scheme provides a homogeneous synthetic aperture in the DSCA. **Figure 5** (e) shows that the amplitude from scattering structures in the image decreases with depth for z > 20 mm. This result can be attributed to the attenuation of ultrasound waves in the medium. For z< 20 mm, the increase in the amplitude with z may be attributed to the spatial response of the array before the focal distance that was not modeled in the reconstruction algorithm.

Table 2. Full width at half maximum (FWHM) along the three directions of the coordinate system for microspheres of Ph1. Values are median ± IQR. Two acquisition modes are compared: a rotate-translate scan at $\Delta\alpha_2$ (dataset of 52 microspheres) and a translation-only scan (dataset of 16 microspheres). The translation range was L=30 mm for both acquisition modes.

| FWHM | Axis of interest | Rotate-translate scan | Translate-only scan |
|---|---|---|---|
| FWHMx | x-axis | 371μm ± 36μm | 1140μm ± 222μm |
| FWHMy | y-axis | 756 μm ± 40μm | 709μm ± 45μm |
| FWHMz | z-axis | 401μm ± 23μm | 373μm ± 26μm |

The FWHM values were assessed for the four rotate-translate scans presented in **Table 1**. **Figure 6** shows that the FWHMy and the FWHMz are stable for all the angular sampling periods $\Delta\alpha$. The FWHMx is significantly larger for $\Delta\alpha_1$, and is on the order of 360 μm (median) for the other angular samplings, with a similar IQR. This result indicates that the angular sampling period $\Delta\alpha_1$ may be too loose to avoid strong side lobes in the vicinity of the main lobe. The steered transmit induced fewer outliers (red crosses), but did not have a significant influence on the median value of the FWHM.

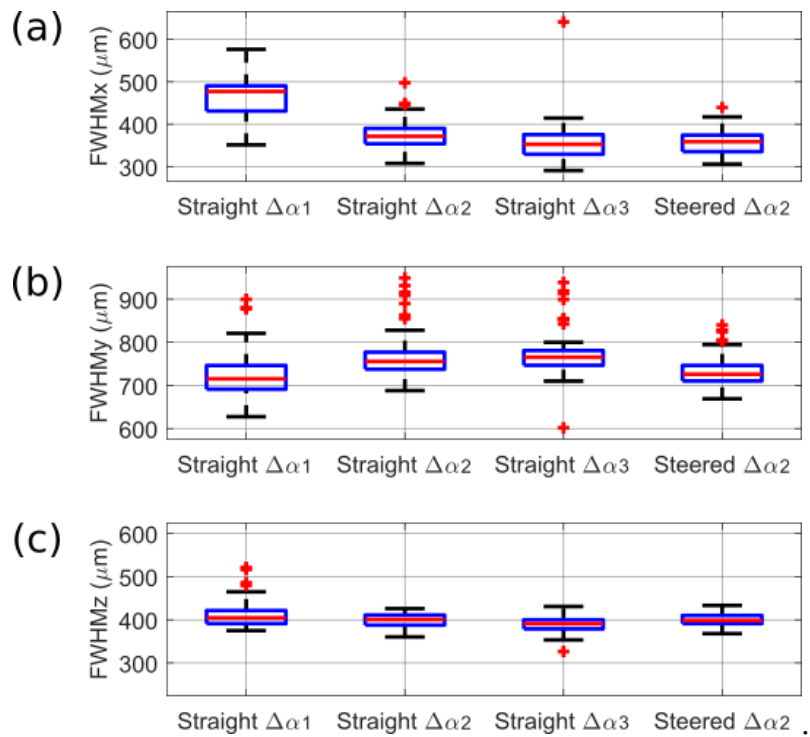

**Figure 6.** Boxplots of the FWHM along the three axes of the coordinate system: (**a**) FWHMx, (**b**) FWHMy, (**c**) FWHMz. The median value is shown with the red line. The box indicates the IQR values and the outliers are shown with red crosses. In abscissa, the scanning name corresponds to the type of transmit and the angular period. The angular periods are $\Delta\alpha_1 \approx 8°$, $\Delta\alpha_2 \approx 4°$ and $\Delta\alpha_3 \approx 2°$.

## 3.3. The cystic resolution (CyR)

The cystic resolution further characterizes the influence of the angular sampling on the image quality by quantifying the effect of the side lobes of the PSF on the image contrast. **Figure 7**(a) presents an image of Ph3 for the plane y = 0 with a logarithmic gray scale ($\Delta\alpha_2$ and $L$ = 30 mm). The side lobes of the LSF are visually similar for the four threads that are all included in the DSCA.

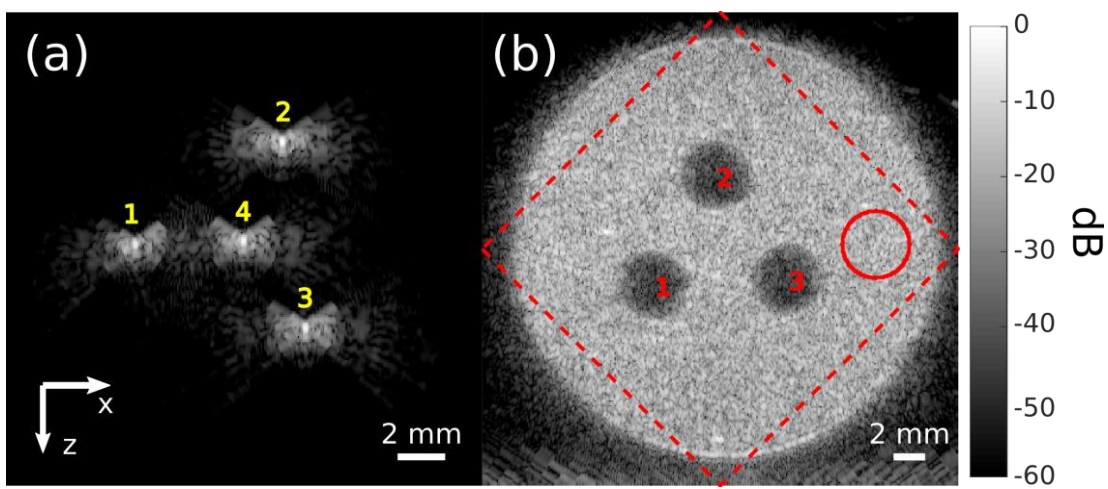

**Figure 7.** (**a**) xz-image of Ph3 (20 mm x 20 mm); (**b**) xz-image of Ph2 (30 mm x 30 mm). Both xz-images are at y = 0 and centered on x = 0 et z = 25 mm. They correspond to acquisitions performed in straight transmit at $\Delta\alpha_2$ with $L$ = 30mm. Numbers identify the threads (**a**) and holes (**b**), respectively. The speckle zone considered to calculate the image contrast is designated with a red circle in (**b**). The images are displayed with -60dB dynamic range.

**Figure 8** (a) displays the cyst contrast $C_{cdB}$ of the image of Thread 2 for the four rotate-translate scans presented in **Table 1**. The iCyR corresponds to a radius that produces a contrast of -20 dB. **Table 3** presents the iCyR values obtained for each thread. The spatial homogeneity of the iCyR can be noticed for a given angular sampling. One exception is observed for Thread 1 and $\Delta\alpha_1$. In this case, the degraded FWHMx resolution for $\Delta\alpha_1$ and the vicinity of Thread 4 biased the iCyR estimation. For straight transmits, the iCyR decreases with the decreasing value of $\Delta\alpha$, which means that the spatial extension of the side lobes decreases and the contrast increases. The difference between $\Delta\alpha_2$ and $\Delta\alpha_3$ indicates that the latter angular sampling period should be preferred in case the image contrast is desired for the targeted application. Interestingly, the iCyR is close for the steered transmit at $\Delta\alpha_2$ than for the straight transmit at $\Delta\alpha_3$. The same result is observed for the $C_{cdB}$ (**Figure 8** (a)). Therefore, globally the lobes of the LSF for $\Delta\alpha_2$ could be mitigated by generating additional tilted planes waves and thereby reinforcing spatial averaging in the image during the reconstruction process.

**Figure 8** (b) compares the anisotropic sectorial cystic resolution aCyR for Thread 2 and the different angular sampling periods. While FWHMz and FWHMx have similar values (**Figure 6**) indicating an isotropic center peak for the LSF, we note that the polar plot of the aCyR has the shape of a lemniscate (∞-symbol) for every angular sampling period. The aCyR is much smaller around the z-axis than along the x-axis. This anisotropy of the side lobes is linked to the finite synthetic angular aperture (around 90°) and its orientation. Therefore, the image contrast depends on the orientation of the structure in the xz-plane. Second, the angular sampling mostly influences the aCyR along the x-axis. A left-right asymmetry of the aCyR can also be noticed. It may be due to the asymmetry of the scanning between the two translation directions in the practical implementation of the scanning scheme (**Figure 3**). On the right side of the polar plot ($\theta \in [-45°; 45°]$), we observe smaller aCyR values for smaller $\Delta\alpha$ in the straight transmit mode. The steered transmit at $\Delta\alpha_2$ leads to aCyR value intermediate between those of the straight transmit at $\Delta\alpha_2$ and $\Delta\alpha_3$.

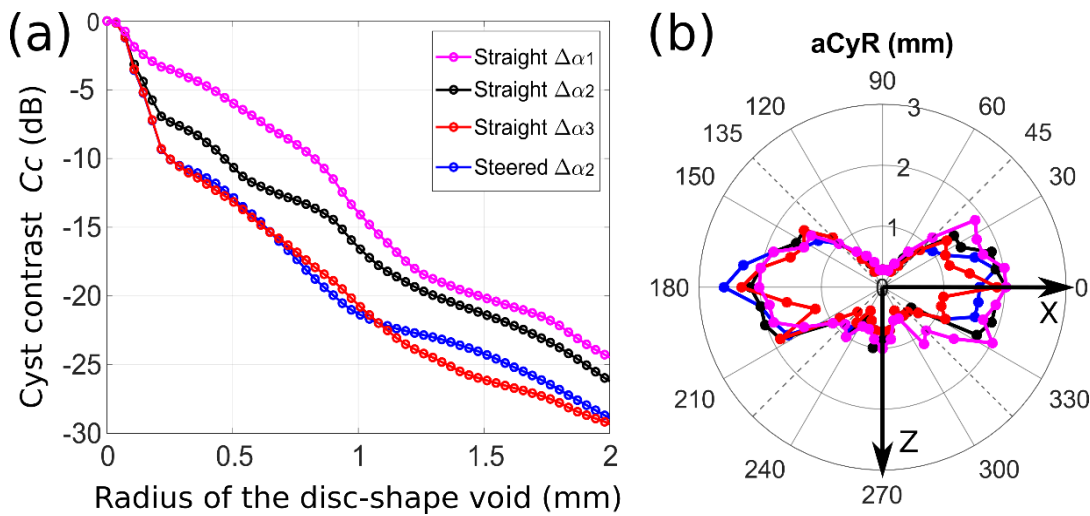

**Figure 8.** (**a**) Cyst contrast in dB, $C_{cdB}$, computed for LSF of Thread 2 as a function of the radius of disk-shaped void considered for equation (6). The curves correspond to the different angular sampling periods and transmit modes; (**b**) polar representation of the anisotropic cystic resolution aCyR assessed on the LSF of Thread 2 for the different angular sampling periods and transmit modes.

**Table 3.** Contrast parameters of the different acquisition modes. Translation range of *L* = 30 mm. The angular periods are $\Delta\alpha_1 \approx 8°$, $\Delta\alpha_2 \approx 4°$ and $\Delta\alpha_3 \approx 2°$.

| Rotate-translate scan | | Straight transmit | | | Steered transmit |
|---|---|---|---|---|---|
| Contrast parameter | Location | $\Delta\alpha_1$ | $\Delta\alpha_2$ | $\Delta\alpha_3$ | $\Delta\alpha_2$ |
| iCyR (mm) | Thread 1 | 2.84 | 1.40 | 1.12 | 1.08 |
|  | Thread 2 | 1.48 | 1.30 | 0.97 | 0.90 |
|  | Thread 3 | 1.51 | 1.37 | 1.08 | 1.08 |
|  | Thread 4 | 1.48 | 1.40 | 1.04 | 0.97 |
|  | *Median* | *1.49* | *1.38* | *1.06* | *1.02* |
| $C_0$ (dB) [1] | Hole 1 | -24.3 | -25.8 | -28.7 | -31.5 |
|  | Hole 2 | -24.4 | -26.2 | -28.9 | -31.1 |
|  | Hole 3 | -24.7 | -25.3 | -27.9 | -31.5 |
|  | *Median* | *-24.4* | *-25.8* | *-28.7* | *-31.5* |

[1] Contrast obtained at the center (i.e. for the smallest cyst radius r = 0.8 mm).

### 3.4. Image contrast

The image contrast was also directly assessed on the three 5mm-diameter anechoic holes of Ph2. **Figure 7**(b) shows an image of Ph2 for the plane y = 0. The holes are identified as well as the speckled background area used for the calculation of $C_0$. The holes 1, 2 and 3 have similar positions to the threads 1, 2 and 3 of Ph3, respectively.

**Table 3** presents the contrast $C_0$ obtained for the four rotate-translate scans presented in **Table 1** and for a cyst area defined by a disk of radius 0.8 mm centered on each hole. First, we note that the cylindrical anechoic holes can all be detected with a contrast higher than 20 dB compared to the speckle background. This result is in agreement with the iCyR values determined in section 3.3, because the radius of the holes is 2.5 mm and the largest median iCyR is 1.5 mm. The $C_0$ values of **Table 3** are similar for the three holes. The contrast can then be considered as spatially homogeneous in the DSCA. However, $C_0$ decreases with the decreasing angular period $\Delta\alpha$ for the straight transmit. Moreover, the contrast is stronger for the steered transmit acquisition at $\Delta\alpha_2$ than for the straight transmit acquisition at $\Delta\alpha_3$. For these two acquisitions, the iCyRs were both on the order of 1 mm. We can note that the disk of the cyst area is at a distance larger than 1 mm from the edges of the hole. As a consequence, the influence of the LSF on $C_0$ might be considered low, and the lower value of $C_0$ for the steered transmit at $\Delta\alpha_2$ is most likely linked to the higher number of tomographic positions in this mode (**Table 1**) and the resulting increased spatial averaging of the electronic noise during the reconstruction process. The $C_0$ values of **Table 3** can be considered as a measure of the contrast to noise of the different acquisition parameters. They all decrease with the number of tomographic positions.

The contrast was evaluated for non-overlapping concentric rings of the same area as the central disk. **Figure 9**(a) displays $C_0$ as a function of the external radius of each ring for hole 1. As expected, $C_0$ reaches 0dB for rings located outside of the hole. Inside and in the vicinity of the scattering edges of holes, $C_{0dB}$ is influenced by the spread of the LSF. As for $C_c$ shown in **Figure 8** (a), the contrast decreases with the decreasing $\Delta\alpha$ for the straight transmit. However, whereas the $C_c$ values were similar for the straight transmit at $\Delta\alpha_3$ and the steered transmit at $\Delta\alpha_2$, $C_0$ is lower for the steered transmit acquisition from a distance of about 1mm from the edges.

**Figure 9** (b)-(e) illustrates the visibility of the anechoic hole 1 for the different rotate-translate acquisitions. For straight transmits, the center of the hole is indeed darker when $\Delta\alpha$ decreases. Additionally, the hole appears larger in the lateral direction (along the x-axis). This result corresponds to the anisotropy of the LSF, identified by the aCyR (**Figure 8**(b)).

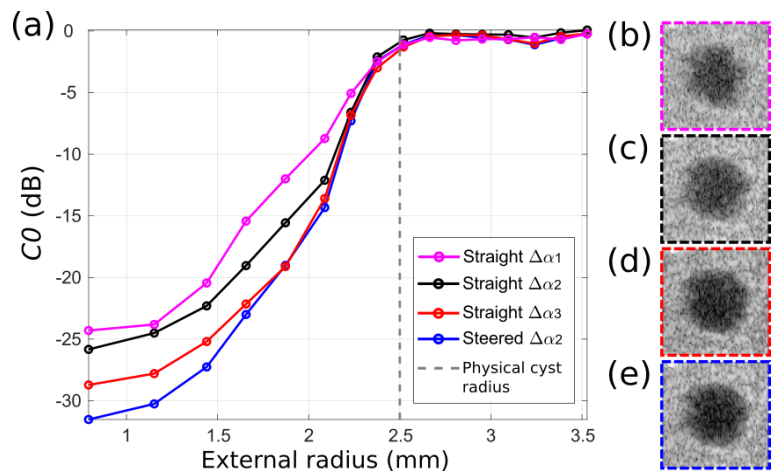

**Figure 9.** (**a**) $C0_{dB}$ curves plotted as a function of the external radius of each ring for Hole 1 and for different angular sampling periods. The angular periods are $\Delta\alpha_1 \approx 8°$, $\Delta\alpha_2 \approx 4°$ and $\Delta\alpha_3 \approx 2°$; (**b**)-(**e**) xz-images of Hole 1 at y = 0 for the different acquisition modes: (**b**) straight $\Delta\alpha_1$, (**c**) straight $\Delta\alpha_2$, (**d**) straight $\Delta\alpha_3$ and (**e**) steered $\Delta\alpha_2$. The images are displayed with -60dB dynamic range.

## 3.5. Complex 3D phantom

To focus on the xz-plane where the rotate-translate scanning scheme has the greatest impact, the scattering structures were confined in a plane for Ph1 and, they were invariant along the y-axis for Ph2 and Ph3. Conversely, Ph4 is a 3-D scattering structure that is not confined to the plane y = 0 containing structures aligned along multiple orientations (**Figure 10** (a)). Threads are arranged in a net with knots and 4-way junctions. The mesh was large to ensure that the branches could be resolved in the rotate-translate mode regardless of their orientation.

**Figure 10** (a) shows a photographic top-view of the net and **Figure 10** (b) displays the MAP US image along the z-axis for a steered transmit at $\Delta\alpha_2$. The two images correspond well, and both the knots and the branches can be identified in the MAP US image. The threads appear continuous even if a local amplitude heterogeneity is visible along the threads with orientations having a significant x-component. This may be linked to randomly distributed microbubbles trapped on the thread. For the junctions, mostly oriented along the y-axis, the lower FWHMy may smooth the reconstruction. The continuity of the threads is also visible on the MAP images along the y-axis (**Figure 10** (d)) and the x-axis (**Figure 10** (e)). As for **Figure 5**, the amplitude decreases with the increasing z coordinate, but the amplitude does not strongly depend on the orientation of the thread.

**Figure 10** (c) and (f) present MAP images corresponding to a translate-only acquisition performed for Ph4 just after the rotate translate acquisition. In the translation-only mode, the resolution along the x axis is degraded throughout the entire imaged volume. This result was expected from 3.2, and it results in smearing and overlapping between the images of some threads (blue arrow in **Figure 10** (c)). The threads also appear discontinuous with bright and dark spots in the translation mode (pink arrows in **Figure 10** (c) and (f)). Video S2 further demonstrates the higher 3D image quality with the rotate-translate mode compared to the translate-only mode.

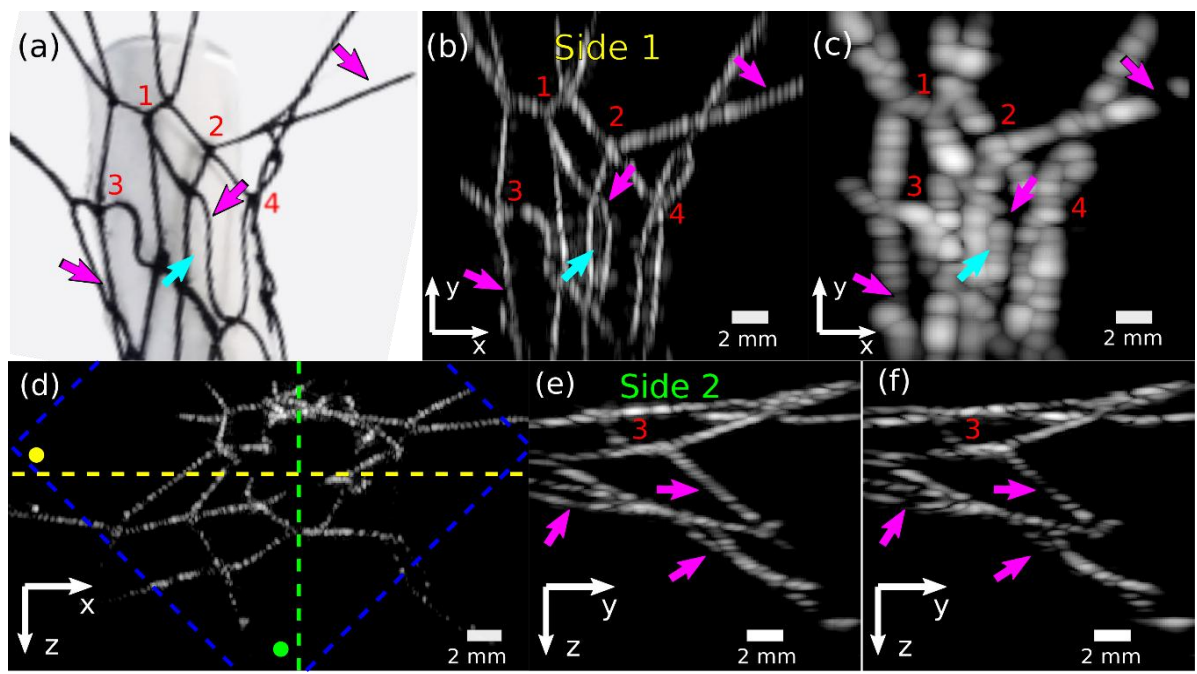

**Figure 10.** Images of a 3-D net. (**a**) Optical picture of the top of Ph4. A paper cylinder was inserted in the net to mask the lower part; (**b**), (**d**) and (**e**) display a volumetric image of Ph4 for a rotate-translate scan at $\Delta\alpha_2$ with a steered transmit; (**b**) MAP image along z-axis of the top part of the volumetric image (part of the volume with z-coordinates smaller than the yellow dotted line in (**d**)); (**d**) MAP image along y-axis; (**e**) MAP image along x-axis of the left part of the volume (part of the volume with x-coordinates smaller than the green dotted line in (**d**)). The voxels outside of the DSCA were discarded for the MAP for (b) and (e); (**c**) and (**f**) display images extracted from the volumetric rendering of Ph4 for a translation-only scan, (**c**) MAP image along z-axis for the same volume as (**b**); (**f**) MAP image along x-axis for the same volume as (**e**); The colorscale for (b)-(f) is between -30 and 0 dB. Knots are identified with numbers that serve as landmarks. Pink arrows indicate threads that appear discontinuous in the image for the translation-only scan and are better mapped with the rotate-translate scan.

## 4. Discussion

We experimentally investigated the performance of a synthetic-aperture rotate-translate approach for volumetric US imaging with a linear ultrasound array. In the elevation direction of the array, this approach demonstrated the synthesis of an angular aperture several times larger than the aperture provided by the native elevation focus of the array. The aperture was increased by a factor of 5 and a Hamming apodization was implemented. As a result, the elevation resolution was improved by a factor 3 compared to a translate scanning approach. Moreover, the aperture synthesis in the elevation plane was obtained over a large area (the DSCA): an area of 4,5 cm² was covered here with a translation range of 3 cm. The elevation resolution was confirmed to be homogeneous in this area. Therefore, the rotate-translate scan was shown to yield superior image quality compared with a translation scan.

To the best of our knowledge, this study is the first demonstration that 3-D ultrasound imaging can be performed with a rotate-translate scanning scheme combined with coherent compounding. A similar scanning approach had previously been proposed for photoacoustic (PA) tomography [20]. However, the ultrasound sources in PA imaging are the illuminated optical absorbers. For PA imaging the transducer is scanned to a different orientation after each laser pulse to receive the ultrasound wavefield generated by absorbers in the medium. For US imaging, the ultrasound waves are both emitted and captured by the scanned US transducer array. As a consequence, the backscattered echoes from different orientations of the exciting US beam need to be recombined coherently to generate a high-quality image. This concept has been extensively used for plane-wave ultrasound imaging [8] and synthetic aperture imaging [31] with a fixed transducer array. However, in this case, the field-of-view and the accessible orientations are limited by the number and geometrical properties of the elements of the array. Coherent compounding was also recently shown for 2-D imaging using multiple transducer arrays with different orientations [32] or a rotation scan of an array [33]. Such approaches increase the angular aperture and the lateral resolution but the transducer arrays were always oriented within the same imaging plane. Similar results were obtained in 2-D PA imaging [34]. Here, we performed 3-D synthetic-aperture scanning by applying coherent compounding from tomographic positions of the array with both different angular elevation orientations and different elevation spatial positions. Additionally, we combined the coherent summing of echoes obtained from different tomographic positions with the summing of echoes from beams electronically steered to different angles. For each steered angle, a dataset with the same translation and angular sampling periods was recorded. The steered angles were chosen to ensure uncorrelated transmitted wave-vectors. As a consequence, the coherent summation over all the steered angles during the reconstruction process reinforced coherent echoes and reduced incoherent noise which led to image contrast enhancement.

The proposed rotate-translate scanning scheme is novel for 3-D US imaging. Furthermore, since both PA and US imaging use the same detector, acquisition parameters optimized for one imaging modality can be applied directly to the other. We benefited from previous innovations in rotate-translate PA tomography such as continuous scanning [27] that showed a reduction in the scan duration. We further reduced the scan duration by more fully assessing the needs in terms of angular sampling. Two main improvements were made to angular scanning: 1) the angular spacing was chosen based on the elevation focusing of the array and 2) a spatial density taper was applied. The smallest angular period considered here $\Delta\alpha_3$ is 3 times larger than the angular period in [20]: $\lambda_c/F = 0,7°$. Therefore, the optimization of the angular spacing reduces the number of tomographic positions by at least a factor of 3. Additionally, the spatial density taper reduced the number of tomographic positions by about 40% and reduced side lobes of the PSF in the elevation plane without any additional amplitude apodization. Indeed, strong side lobes are expected at ± 45° with no apodization, but due to the spatial density taper used here they were not stronger at 45° than at 0° (**Figure 8** (b)). Finally, the total number of tomographic positions was thus reduced by at least by a factor 5 (80% decrease) relative to the number previously reported in [20].

Three different angular periods $\Delta\alpha_1 \approx 8°$, $\Delta\alpha_2 \approx 4°$ and $\Delta\alpha_3 \approx 2°$ were tested. The number of tomographic positions and the scan duration increase by a factor two when the angular period is divided by two (**Table 1**). Metrics were computed to evaluate the impact of the angular sampling period on the image quality. We found that $\Delta\alpha_1$ leads to a significant degradation of the FWHM resolution along the x-axis compared to $\Delta\alpha_2$ and $\Delta\alpha_3$. Therefore, $\Delta\alpha_1$ was determined to be too loose to prevent side lobes from degrading the main lobe of the PSF. The cystic resolution and the contrast on the edge of an anechoic region were also improved with the decreasing angular period. For the angular sampling $\Delta\alpha_2$, we performed acquisitions in straight transmit mode and in steered transmit mode. Interestingly, both contrast measurements improved with the addition of steered angles, while the angular sampling was the same. We can explain this result by the enhanced spatial averaging of the "diffraction noise" when steered transmits are performed. Moreover, the contrast parameters ($Cc$ and $C0$) were found to be similar with the angular sampling $\Delta\alpha_2$ in steered transmit and with $\Delta\alpha_3$ in straight transmit. The steered transmit does not modify the acquisition duration but does substantially increase the size of the dataset, while the scan duration at $\Delta\alpha_2$ is half that at $\Delta\alpha_3$. Therefore, a gain in the image quality can effectively be made without compromising the acquisition duration by the addition of a steered transmit. A systematic evaluation of the contrast for different numbers of steered angles is beyond the scope of this paper, but is of great interest for our rotate-translate system and will be considered in future studies. The angular periods $\Delta\alpha_2$ and $\Delta\alpha_3$ will also be implemented in PA tomography. As the tomographic system is implemented with a programmable ultrasound machine that can record US waves upon the arrival of a trigger pulse, PA tomography sequences with a pulsed laser could be implemented during scans. The interlacing of PA and US acquisitions will be investigated in future studies for simultaneous PA and US 3D imaging of the same volume with the scanner. Additional strategies such as tissue harmonic coherent compound imaging using advanced US sequences to isolate non-linear echoes [35, 36] and the use of contrast agents could be integrated within the tomographic system in the future to further improve US image contrast and specificity. Because of the scan duration, it is expected that contrast agents that could be retained within biological tissue would work best such as targeted microbubbles [37] or nanodroplets converted to microbubbles upon optical [38,39] or ultrasound [40] activation. Feasibility has been shown, for example, to optically activate microbubbles from nanodroplets that are detectable with PA imaging so that, with the choice of appropriate dual-modality contrast agents PA and US can be used together to characterize agent arrival and delivery [39].

The system was developed with a single-side access for an accessibility to various body parts. The scan duration is also a crucial point for *in vivo* application, clinical or preclinical. Long acquisitions may cause discomfort and tissue motion that can cause image blurring. As mentioned above, the number of tomographic positions was already reduced by adjusting the spatial sampling. But the acquisition duration is the number of tomographic positions multiplied by the acquisition rate. With the ultrasound machine used in this study, the acquisition rate could be in the kHz range. Currently, we limited the acquisition rate to 50 Hz because of the maximum acceleration of the rotation stage (see section 2.3.2). The acquisition duration could then be greatly reduced with a faster rotation stage such as one with a magnetic motor. With our worm gear motor, we achieved a scan duration of 10 seconds for a 9 $cm^3$ volume. This acquisition time is smaller compared to other rotate-translate scanning schemes that have been applied *in vivo* in PA tomography [41] and for incoherent spatial compounding of Doppler data [42,43]. Although these studies map functional information and not only pulse-echo US imaging, they give a reference for *in vivo* acquisition duration. Li et al [41] performed multispectral PA tomographic acquisitions in 12 minutes. For 3D Doppler tomography with incoherent compounding of Doppler images, reported acquisition times were of 15-20 min [42,43]. To avoid blurring due to respiratory motion, acquisition methods such as respiratory gating, i.e. acquisition only during the respiratory pause, can be considered. A weak oversampling can be applied to compensate for missing tomographic positions. With a current acquisition duration of 10 s for a 9 $cm^3$ imaged volume (in the DSCA) in the steered transmit mode and a reconstruction duration of 95 s (**Table 1**), the scanning system can currently achieve a volumetric imaging rate of 0.57 volumes/min. The reconstruction duration could be further accelerated with a high-end GPU unit. However, even if the acquisition rate between tomographic positions was set in the kHz range, which is the maximum rate given the ultrasound propagation time and the ultrasound hardware, the full-volume acquisition duration will still be on the order of one second in the steered transmit mode. Therefore, real-time volumetric imaging (> 5 volume/s [4]) would only be achievable with degraded image quality (rotate-translate scan in the straight transmit mode and with $\Delta \alpha \geq \Delta \alpha_2$ or a translate-only scan). Such an approach could be used to position the volume of interest in the DSCA with a real-time display prior to a higher quality image acquisition.

Finally, the developed scanning method is applicable to a broad range of already-available US arrays. The scan parameters best-adapted for each array can be calculated based on its geometrical properties and the center frequency. Moreover, the one-side access design and the home-made (3-D printed) holder make the system easily adaptable to any kind of available probes. Although the feasibility was demonstrated with a 5-MHz ultrasound array, the method is of great interest for higher frequencies within the range of medical ultrasound (> 15 MHz), in particular to improve the sensitivity to smaller tissue scatterers and the resolution compared to low frequencies. For applications and ultrasound acquisition modes that do not require high frame rate imaging, we believe that our 3-D synthetic-aperture rotate-translate scanning method could greatly enhance the capacity of 3D biomedical ultrasound imaging without the complex development of new transducer arrays. Among clinical applications for which 3D ultrasonography has already been shown to be of high interest, our approach could specifically benefit those requiring a high contrast and resolution such as quantification of carotid atherosclerosis [44] and use of musculoskeletal ultrasonography to assess rheumatoid arthritis [4,45]. Additionally, we expect that our 3D scanning method will help longitudinal monitoring of pathological tissues for which high quality volumetric images are crucial for image processing and for accurate comparison of images acquired at different points in time.

## 5. Conclusions

We demonstrated here that a clinical ultrasound linear array designed for 2-D imaging can be used to obtain high resolution and contrast volumetric US images with a 3-D synthetic-aperture scanning method. We developed and experimentally implemented an advanced scanning scheme combining rotation and translation motions of the array. The angular sampling period was adjusted and tapered to optimize the scan duration without compromising the image quality. An acquisition duration of 10 s was achieved for a volume of 9 cm$^3$. The image contrast was further enhanced by the addition of electronically steered exciting beams. The method is expected to benefit applications that require high image quality but a relatively low volumetric scanning rate such as longitudinal monitoring of pathological tissues.

**Supplementary Materials:** Video S1: Video of the experimental set-up showing the motion of the stages for an acquisition performed at $\Delta\alpha_2 \approx 4°$ with $L$ = 30mm. Video S2: 3-D volume images of the net presented in **Figure 10** compared for the scan performed using the rotate-translate geometry (left) and the scan performed with translation only (right). Rotating maximum amplitude projection images around an arbitrarily selected z axis are displayed with a 10 deg angle between the projections. The grayscale display is between -30 and 0 dB for both images. The image display was obtained with the 3-D project option of the software ImageJ [46]. Figure S1: US images of Ph 1 performed with different translation ranges $L$. (**a**) Optical picture showing the spatial distribution of microspheres. (**b**) to (**d**) Maximum-amplitude projection (MAP) images along the y-axis for $L$ = 30, 20 and 10 mm, respectively. This figure displays the full-size MAP images of **Figure 4** (c) and (d).

**Author Contributions:** Conceptualization, J.G.; methodology and investigation, T.L., I.Q and J.G.; software, T.L., I.Q and J.G.; validation, T.L., I.Q., L.B. and J.G.; .formal analysis, I.Q. and J.G.; writing—original draft preparation, T.L., I.Q and J.G; writing—review and editing, L.B.; supervision, L.B.; project administration, J.G.; funding acquisition, J.G. All authors have read and agreed to the published version of the manuscript.

**Funding:** This research was funded by the MITI CNRS (Defi Imag'IN 2017-2018), Gefluc Paris- Ile de France (2017) and Sorbonne Université (Emergence 2019-2020). This work was performed by a laboratory member of France Life Imaging network (grant ANR-11-INBS-0006).

**Acknowledgments:** Isabelle Quidu thanks the Université de Bretagne Occidentale for their support to the 12-month visiting stay (Congés pour Recherches ou Conversions Thématiques) at the Laboratoire d'Imagerie Biomédicale from September 2019. The authors thank the Life Imaging Facility of Paris Descartes University (Plateforme Imageries du Vivant – PIV) for their technical support and fruitful discussions. The authors thank Pascal Dargent (Laboratoire d'Imagerie Biomédicale, LIB) for the mechanical design of some scanner elements, and Laurent Fabre (Institut des Systèmes Intelligents et de Robotique, ISIR) for machining the elements. The authors thank Marwa Chammakhi for her work at the early stage of the system development.

**Conflicts of Interest:** The authors declare no conflict of interest. The funders had no role in the design of the study; in the collection, analyses, or interpretation of data; in the writing of the manuscript, or in the decision to publish the results.

# Appendix A

In appendix A, we give details about the reconstruction algorithm. In particular, how we computed: 1) the two-way travel times between the US transducers and the center of each imaged voxel and 2) the apodization windows.

The 3-D image voxel grid is defined in the fixed coordinate system (O, $\mathbf{e_x}$, $\mathbf{e_y}$, $\mathbf{e_z}$) presented in **Figure 1**(b). In the xz-plane, the origin O corresponds to the rotation axis when $l_{trans}$ = 0. Along the y-axis, O corresponds to the center of the active array aperture. The mobile Cartesian coordinate system ($O_a$, $\mathbf{u}$, $\mathbf{v}$ $\mathbf{w}$) is attached to the transducer array (**Figure 1**(a)). In 3-D, the origin $O_a$ is set to the center of the array, and the elements of the array are uniformly distributed along the v-axis. For computation of the ultrasound pulse travel times, transformation from ($O_a$,$\mathbf{u}$,$\mathbf{v}$,$\mathbf{w}$) to (O, $\mathbf{e_x}$, $\mathbf{e_y}$, $\mathbf{e_z}$) is performed in two steps. First, the coordinates of the basis ($\mathbf{u}$,$\mathbf{v}$,$\mathbf{w}$) are computed by taking into account potential misalignments of $\mathbf{v}$ with $\mathbf{e_y}$ and the rotation angle $\alpha_{rot}$. Then, the coordinates of $O_a$ in (O, $\mathbf{e_x}$, $\mathbf{e_y}$, $\mathbf{e_z}$) are computed given a possible mismatch between $O_a$ and O when $l_{trans}$ = 0 and $\alpha_{rot}$ = 0, the rotation angle $\alpha_{rot}$, and the signed translation length $l_{trans}$. Misalignment parameters were determined with a calibration process: a known object (thread) was imaged and parameters were adjusted to maximize the amplitude of the reconstructed image.

For the pathway from the US array to the voxel, we assume that cylindrical waves are emitted. The distance between the emitting axis and the center P of one voxel is computed and divided by the speed of sound to get the travel time. The location of the emitting axis depends on the tomographic position and the steering angle $\gamma_{st}$. For the straight transmit ($\gamma_{st}$=0), the emitting axis is the v-axis. For $\gamma_{st} \neq 0$, the first firing element (at one edge of the emitting aperture) stays on the emitting axis, but the direction of the axis is rotated by $\gamma_{st}$ around $\mathbf{u}$. The u-coordinate of P in the mobile coordinate system is computed and compared with $d_u$ (**Figure 1** (b)) to implement the apodization detailed in section 2.2. This apodization accounts for the focusing of the transducer elements in the elevation direction (u-axis) by limiting the contribution of the signals to a slice of width 2*$d_u$, which is taken to be equal to the pulse-echo -12 dB full width of the ultrasound beam [21] (section 2.2.). This slice width is then related to the angular aperture β by the equation $d_u$ = 0.3* $\lambda_c$ / tan(β). For the pathway from the voxel to the US array, we assume that the echo is a spherical wave emitted by P and detected by point detectors located at the center of each element of the array.

For each tomographic position, a dynamic aperture approach was implemented to keep the angle of acceptance in the lateral direction (along the v-axis) as constant as possible and thereby obtain a homogenous resolution in the $e_y$-direction for all the points in the 3D image [47]. A lateral f-number of 1.3 was considered. It corresponds to the ratio between the elevation focus of the array F and the reception aperture (64 elements), and to a compromise between the resolution value and its homogeneity. This dynamic aperture was apodized with a Hamming window to avoid strong side lobes.

From all the tomographic positions of the transducer during the experimental implementation of scanning, we discarded positions outside of the translation limits of the theoretical scanning scheme, such as from B to C in **Figure 3**. For the remaining positions (**Figure 3**), we applied weights equal to the actual velocity divided by the targeted velocity of equation (3). This apodization avoids higher amplitudes on the edge of the imaged area (**Figure 1**(c)).

The reconstruction algorithm was coded in CUDA and executed on GPU.

## Appendix B

In appendix B, we explain the theoretical basis for choosing the angular sampling periods $\Delta\alpha$ for the rotate-translate scan. The translation period for the rotate-translate scan was set to $\Delta l = H$ as discussed in section 2.2 Additionally, we justify the translation sampling period chosen for the benchmark translation-only scan.

For the rotate-translate scan, the angular sampling problem is related to two different configurations: sampling for a circular ring array [22] and linear sampling with finite size transducers [24]. First, Simonetti *et al* [22] derived a spatial sampling criterion for imaging objects within a circular array comprised of point transducers deployed over the entire circular aperture. In such a case, the angular sampling period $\Delta\alpha$ depends on the maximal radial dimension $r$ of the imaged object and should be adjusted so that $\Delta\alpha < \lambda_c / (2*r)$ to reject grating lobes outside of the imaged area. $\lambda_c$ is the wavelength at the center frequency of the transducer. There is no fixed center of rotation in the coordinate system (O, $e_x$, $e_y$, $e_z$) in our synthetic rotation array, however we may consider that the rotation occurs around each point in the imaged area and, due to the amplitude masks in the reconstruction process, $r$ could be replaced by $d_u$ = 0.6 * $\lambda_c$ * f-number. In this case, the sampling criterion would be $\Delta\alpha < 1/(1.2 *$ f-number). This solution is independent of $\lambda_c$. Second, Cox *et al* [24] note that the finite size of a transducer element acts as a low-pass spatial filter reducing the Nyquist-Shannon frequency compared to point detectors. For a synthetic planar array comprised of disk transducers of radius $a$, Cox *et al* showed that an analysis in the spatial Fourier space leads to a sampling frequency equal to $a/4$, as opposed to the sampling frequency of $\lambda_c /2$ that would be required for point detectors [24]. We apply similar reasoning. The angular aperture of the transducer is [-β, β] with β = atan($D/(2*F)$) ≈ 1/(2 * f-number)  (**Figure 1** (a)). Therefore, in the Fourier space, the maximum angular frequency can be considered to be equal to 1/(2.β). Thus, the minimum angular sampling frequency can be set to 1/β, and $\Delta\alpha \leq \beta$. This solution is also independent of $\lambda_c$ and, it is more restrictive than the criterion based on the circular array. However, in order to be sure to discard any aliasing effects, the maximum angular frequency can be considered as equal to 1/β or 2/β. These frequencies lead to angular sampling periods equal to $\Delta\alpha$ = β/2 and $\Delta\alpha$ = β/4, respectively.

For the scans using only a translation, we applied the criterion based on the analysis developed for linear sampling with finite size transducers [24]. For our focused transducer, we considered that the transducer could be modeled as a segment of length *H* located at the focus where *H* is the two-way FWHM of the ultrasound beam. Therefore, an appropriate sampling period for translation is $\Delta l = H/8$ [24].